\documentclass[a4paper]{article}
\usepackage{RR-Inria/RR}
\usepackage{hyperref}
\usepackage{algorithm2e}
\usepackage{graphicx}
\usepackage{amsfonts}
\usepackage{epsf}\epsfverbosetrue
\usepackage{alltt}
\usepackage{subfigure}
\usepackage{color}
\usepackage{float}
\usepackage{url}

\usepackage[T1]{fontenc}
\usepackage{ulem}
\usepackage[final]{changes} % use [final] instead of [draft]
\usepackage{amsmath,latexsym}
\usepackage{times}
\usepackage{cite}

\usepackage{graphicx,epsfig} % Si le document doit inclure des images
\usepackage{listings}
\usepackage{color}
\usepackage{epsf}\epsfverbosetrue
\usepackage{alltt}
\usepackage{subfigure}

\usepackage{algpseudocode}
\usepackage{ioa_code}

\usepackage{boxedminipage, epsfig,graphics,graphicx,subfigure, amsmath,amssymb,multirow, array}
\usepackage{newalg}
\usepackage{url}
\usepackage{color}

\def\BibTeX{{\rm B\kern-.05em{\sc i\kern-.025em b}\kern-.08em
    T\kern-.1667em\lower.7ex\hbox{E}\kern-.125emX}}

\newcommand{\remove}[1]{}

\newcommand{\psfigure}[3]{ % {scale}{filename=label}{caption}
 \begin{figure}[\placement]\begin{center}%
  \epsfig{file=figs/#2,width=#1\hsize}%
\let\normalsize\small\caption{#3\label{fig:#2}}%
  \end{center}
\end{figure}}

\definechangesauthor{FLF}{blue}
\definechangesauthor{AV}{green}
\definechangesauthor{CS}{red}
\definechangesauthor{AP}{magenta}

\RRdate{May 2010} %%
\RRNo{7285}

\def\email#1{~}

\RRauthor{
 Mubashir Husain Rehmani\thanks{Lip6/Universite Pierre et Marie Curie - Paris 6, France} 
 \and Aline Carneiro Viana\thanks{INRIA, France and TU-Berlin, Germany}
 \and Hicham Khalife\thanks{LaBRI/ENSEIRB, France}
 \and Serge Fdida\thanks{Lip6/Universite Pierre et Marie Curie - Paris 6, France}
 }

\authorhead{Husain Rehmani \& Carneiro Viana \& Khalife \& Fdida}
\RRtitle{Une architecture d'accès à internet basée sur la radio
cognitive pour le déploiement des réseaux en cas de catastrophes dans des environnements hostiles \\}

\RRetitle{A Cognitive Radio Based Internet Access Framework for Disaster Response Network Deployment} %%
\titlehead{Internet Access Framework for Disaster Response Network Deployment}
\RRresume{Dans cet article, nous proposons une architecture d'accès à internet basée sur la radio
cognitive pour le déploiement des réseaux en cas de catastrophes dans des environnements hostiles. L'architecture proposée est conçue pour aider les réseaux existants et particulièrement ceux endommagés pour reconstituer leur connectivité et les relier à l'Internet. L'architecture proposée constitue la base pour le développement de protocoles et  algorithmes pour les réseaux radio cognitifs déployés dans des \\
environnement hostiles.

}
\RRabstract{
In this paper, we propose a cognitive radio based
Internet access framework for disaster response network deployment
in challenged environments. The proposed architectural
framework is designed to help the existent but partially damaged
networks to restore their connectivity and to connect them to the
global Internet. This architectural framework provides the basis
to develop algorithms and protocols for the future cognitive radio
network deployments in challenged environments.

}
\RRmotcle{Réseaux radio cognitifs, environnement hostiles, architecture d'accès à internet.}

\RRkeyword{Cognitive radio networks, challenged environments, Internet access framework, disaster response networks.} %%
\RRprojets{Asap}
\RRtheme{\THCom}
\RCSaclay % Sophia Antipolis M\'editerran\'ee

\begin{document}
\normalem
\makeRR   % cas d'un rapport de recherche

%\if@draftclsmode% draft mode uses larger than normal spacing
\def\baselinestretch{1.5} % controls line spacing for draft version
%\fi

\section{Introduction}
\label{sec:in}

Natural disasters like earthquake or storms are unpredictable and rather a frequent phenomenon these days. The collapse of communications infrastructure
is the usual effect of disaster. In fact, different types of communication networks could be affected by a disaster. For instance, base stations of cellular networks, lack of connectivity between sensors and sink in static WSNs, damage in the existing WLANs etc. Thus, these partially damaged coexistent networks that were previously deployed are now disconnected and all kind of wireless communication cease to work.

In spite of technological advancements, the instantaneous deployment of core telecommunication infrastructure, for e.g. base stations in the case of cellular networks,  is not feasible because of planning and cost. Besides, there is a quick need to help rescue teams or NGOs to facilitate organized help and rehabilitation works, which further motivates the need of rapid ad-hoc network infrastructure deployment. {\it The goal of this rapidly deployed ad-hoc network infrastructure is to provide connectivity and Internet access to partially destroyed networks and to help the rescue team members, until the telecommunication infrastructure is repaired}. The use of WLANs can be the solution, and that is in fact what in Hiati~\cite{hiati} engineers did but the achieved communications suffered considerable delays. However, these disaster response networks, which we referred as challenged networks, impose several constraints like intermittent connectivity, delay, high error rates, no end-to-end paths, unreliable links, heterogeneous devices and operating environment, lack of infrastructure, to name a few. In fact, disasters and emergencies are unpredictable, and network deployment should allow rapid and ad-hoc deployment, and must be specifically designed to cater the needs of challenged environments.

In this paper,  {\it we propose a Cognitive Radio Based Internet Access Framework for Disaster Response Network Deployment in Challenged Environments}. In fact, cognitive radio is the term first coined by J. Mitola {\it et al.}~\cite{mitola} in the year 1999 for efficient spectrum utilization. These cognitive radio devices are capable to change their operating parameters on-the-fly and adapt with respect to the environment~\cite{crahns}. By doing so, Cognitive Radio (CR) devices opportunistically exploits the spectrum left unoccupied by Primary Radio (PR) nodes. Moreover, cognitive radio devices operate in an ad-hoc fashion and form cognitive radio ad-hoc networks. In this way, through our proposed framework and by exploiting cognitive radio technology, the goal of providing robust connectivity and Internet access to partially destroyed networks can be easily achieved.  In this context, to allow CR devices to restore the connectivity of partially destroyed coexistent network as well as providing Internet accessibility, an architectural framework is required, which in turn provides rapid, cost-effective, and robust connectivity. Recently,~\cite{gorcin} discussed the use of cognitive radio in public safety systems, while a cognitive agent based approach for post-disaster communication is proposed in~\cite{sonia}, nevertheless, the connectivity of partially destroyed networks and their connectivity to the global Internet is still an unaddressed topic.

In challenged environments, {\it Cognitive Radio Ad-Hoc Networks (CRNs) is a promising technology and capable to federate the communication of coexistent networks temporarily.}
In fact, several distinguished features of cognitive radio technology make CRN an easy to deploy and flexible solution for challenged environments. These features include for instance the accessibility and flexibility of communication over the whole spectrum band.
%This feature can be used in tsunami like catastrophe to provide fast access to radio spectrum until the rescue team arrives~\cite{mitola1}.
Another important feature is the multi-radio capability of CRNs which can further be used to control communication overhead. For instance, if one radio is available per device, flooding should be avoided in order to limit communication overhead, what requires the design of different forwarding mechanisms. However, if multi-radios are available, flooding can become a powerful transmission strategy, if access to channels is well done.

%\ali{[we are doing flooding with SURF. here add that, otherwise, if multi-radios are available, flooding can become a powerful transmission strategy, if access to channels is well done, something like that.]}

Despite the availability of whole communication spectrum band, concentration of all the radio devices over a single spectrum band %\footnote{Traditional wireless ad-hoc networks operates in ISM band and wireless devices specific to these networks have limited available spectrum.} 
could lead to contention and collision problems, which further reduces the connectivity to global Internet. Cognitive Radio could help a lot on these, by providing more ``communication space'' to devices. In addition, the inherent self-organizing capabilities of cognitive radio devices i.e. detect and jump among channels, features provided by the spectrum sensing and allocation, and the ad-hoc deployment can also be interesting for post-disaster situations.

The remainder of this paper is organized as follows: Section~\ref{conv} discusses the related work. The architecture of the proposed framework is presented in Section~\ref{sec:iafcrn}. Issues and challenges concerning deployment and connectivity of the proposed framework are discussed in Section~\ref{deploy}. Section~\ref{surf} discusses the use of channel selection strategy SURF in conjunction with the proposed architecture. Finally, we conclude in Section~\ref{sec:conclusion}.

\section{Related Work}

\label{conv}

Recent work on the deployments of cognitive radio networks in post-disaster situations include~\cite{gorcin} and~\cite{sonia}. The authors in~\cite{gorcin} discussed several opportunities from the aspect of cognitive radio in public safety, however, these discussions remained general without proposing a feasible architecture or framework. A cognitive agent based approach for post-disaster communication is proposed in~\cite{sonia} but the authors did not explain how the connectivity of partially destroyed networks as well as their connectivity to the global Internet could be achieved.
In summary, the problem of restoring connectivity of partially destroyed networks and connecting them to global Internet is still unaddressed.

Another cognitive radio based network architecture DIMSUMnet~\cite{dimsumnet} is proposed for the integration of cognitive radio networks into the global Internet. This architecture is specifically designed for cellular networks and based on a centralized regional spectrum broker and does not cater the needs of rapid and ad-hoc network deployments, mostly required in post-disaster situations. Moreover, this architecture requires significant time in planning and deployment, and requires strong coordination with the existing infrastructure, instead of opportunistic, distributed, rapid, and un-coordinated deployments mostly required in challenged environments. Compared to~\cite{dimsumnet}, our proposed architecture is totally rapid, distributed, cost-effective, suitable for ad-hoc deployment and provides robust connectivity to partially destroyed networks.

In addition, cognitive radio ad-hoc networks has been widely used in several application scenarios including military and mission-critical networks~\cite{militarysdr},~\cite{ossama}, and consumer-based applications~\cite{applications},~\cite{buddhikot},~\cite{chapin}. Cognitive radio technology can also play an important role in E-health applications~\cite{ehealth},~\cite{ehealth1}. These aforementioned works are not suitable for post-disaster situation and can not be directly implemented in such scenarios. Consequently, there is a need to exploit inherent features of cognitive radio technology and tailor them to be well operated for disaster response networks. Thus, keeping altogether in mind; the adaptive and inherent self-organizing capabilities of cognitive radio devices~\cite{simon}, their potential applications, and recent advancement in hardware technology~\cite{zhang}, make the cognitive radio technology the most suitable radio access technology for challenged environments. 

\section{An Internet Access Framework for Future Cognitive Radio Networks}
\label{sec:iafcrn}
\begin{figure}[hbtp]
    \begin{center}
    {
        \includegraphics[width=12cm, height=6.5cm]{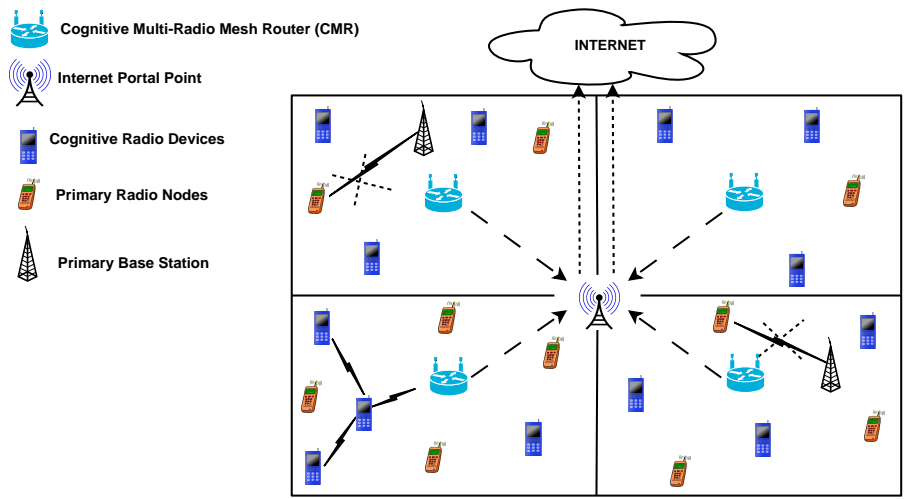}
    }
\caption{An Internet Access Framework for Future Cognitive Radio Networks} 
\label{fig_1}
\end{center}
\end{figure}

Internet Access Framework for Future Cognitive Radio Networks is a three-tier architectural framework tailored to implement and deploy real cognitive radio network applications in challenged communication environments. Fig.~\ref{fig_1} depicts the concept of Internet Access Framework for Future Cognitive Radio Networks. The building blocks of this architecture are: (1) Cognitive Radio (CR) devices, (2) Cognitive Multi-Radio Mesh Routers (CMR), and (3) Internet Portal Point. 

%\ali{[Mubashir, your figure has still AP. the other figures are really nice. could not you make this better too? instead of circles and rectangles use some real entities, for CMR, IPP, CR, etc]} \ali{[this paragraph is better here, because you say a three-tier without describing what are the components...]}

In this architecture, we consider partially destroyed networks as primary networks and their nodes as primary nodes. Indeed, our objective is to detect on-going communications of the partially destroyed infrastructures in order to offer them connectivity to other parts of the same infrastructure or even to the Internet.
%This is primarily because we want somehow to detect the on-going communications that is going on in partially destroyed networks in order to restore their connectivity.
It is clear that interconnecting different type of networks using different technologies can be considered as a challenging task, however, the flexibility and dynamic spectrum management offered by CRN can help to overcome these obstacles. In such kind of situations non-CR devices can be considered as PR nodes, since they are the users of the network with high priority and their communication can not be affected. Non-CR devices need to communicate with the CR devices in order to restore their connectivity to other parts of the network and Internet. If they detect disconnectivity to other non-CR devices, Access Points, sink, base station or Internet, some how the non-CR devices has be have a mechanism to detect the bad performance of their network, then they contact CR devices. Practically, when non-CR devices, need to communicate with CR devices, they first need to detect them. To achieve this goal, CR devices can advertise their presence to non-CR devices. Moreover, CR devices have to overhear the channels in order to know if the data transmitted by a non-CR device is for another non-CR device or for a CR device - to reach the Internet.

%\ali{[first here you have to say, in such kind of situations non-CRs can be considered as PR nodes, since they are the users of the network with high priority and their communication can not be affected...etc. and why non-CRs need to communicate with CRs... you just jump from an explanation to another. explain better. if they detect disconnectivity to other non-CR, AP, sink or internet (some how the non-CRs has be have a mechanism to detect the bad performance of their network), then they contact CRs.]}

%\ali{[agree, but give reasons why do you consider this? other important point here, CRs can not affect the non-CR transmissions, but the some non-CR will need to communicate with CRs, for this, they need to detect than. So, CRs can advertise they presence to non-CRs, and CRs has to overhear the channels in order to know if the data transmitted by a non-CR is to another non-CR or to a CR - to reach the interent. Discuss these issues here.]}

This architecture can be operated in two scenarios: single-hop and multi-hop. CR devices communicate directly with the cognitive multi-radio mesh routers in single-hop scenario, while in multi-hop scenario, CR devices create multi-hop path to reach to the nearest cognitive multi-radio mesh router. 

\subsection{Architecture}
\label{arch}

%Fig.~\ref{fig_4} shows how the proposed architecture is restoring the connectivity of partially destroyed network and help them to connect to the Internet. We now describe individually the functionality of these components:  \\

\begin{figure}[htbp]
    \begin{center}
    {
        \includegraphics[width=12cm, height=6cm]{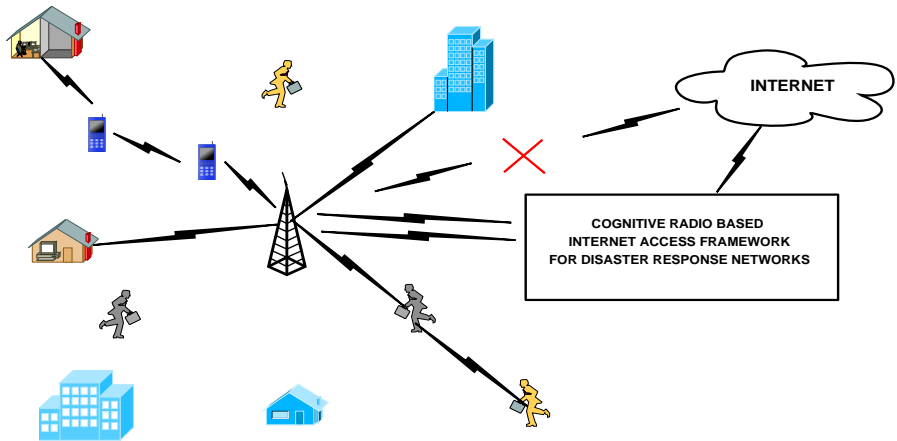} 
    }
\caption{A cognitive radio based disaster response network restore the connectivity of partially destroyed network to the global Internet. }
\label{fig_4}
\end{center}
\end{figure}

In Fig.~\ref{fig_4} we show a practical use case of our framework. In the shown scenario, our architecture acts as a gateway able to federate various existing infrastructures and restore their connectivity to Internet. We now describe individually the functionality of the framework components. 

%\ali{[very nice arguments, but in the figure you should also put some CRs in multi hop. it gives the impression the framework is only for Internet connectivity.]}

\subsubsection{Cognitive Radio (CR) Devices}

Cognitive radio devices which are based on software defined radio can access any cognitive multi-radio mesh router to upload their data to the Internet. These devices can be mobile and are capable to directly communicate with the cognitive multi-radio mesh router in single-hop fashion, they can also create multi-hop network to reach nearest cognitive multi-radio mesh router.

In single-hop scenario, cognitive radio devices are not responsible for their channel selection decision, instead cognitive radio devices will provide feedback to the cognitive multi-radio mesh router about the spectrum occupancy. Cognitive multi-radio mesh router will then make an intelligent decision about channel selection and communicate them to CR devices in order to be used for data transmission.

\begin{figure}[htbp]
    \begin{center}
    {
        \includegraphics[width=12cm, height=5.5cm]{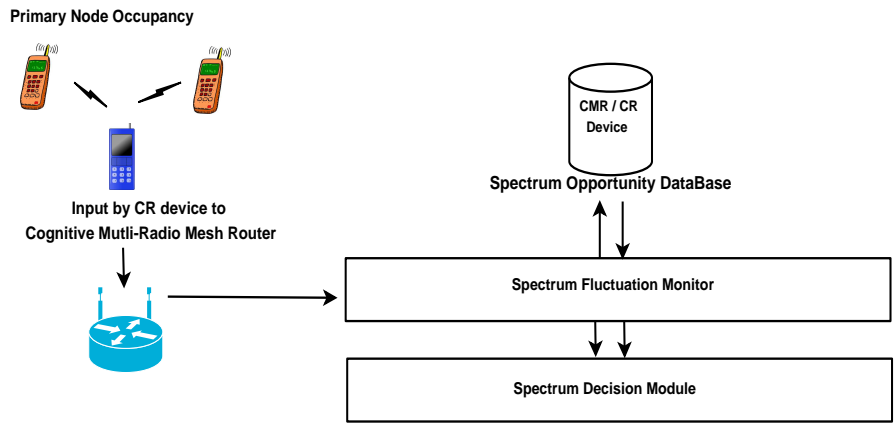} 
    }
\caption{Function of Spectrum Fluctuation Monitor in CMRs. }
\label{fig_3}
\end{center}
\end{figure}

In the multi-hop scenario, the accessibility of CR devices to the cognitive multi-radio mesh router is quite challenging due to lack of any centralized authority. Thus, CR devices are responsible themselves to collect the locally inferred spectrum related information and make a channel selection decision alone to reach the cognitive multi-radio mesh router. Therefore, an intelligent channel selection techniques should be employed that facilitates CR devices in their channel selection decision. In multi-hop scenario, cognitive radio devices should be equipped with spectrum fluctuation monitor to support intelligent spectrum decision (cf. section~\ref{ap}, fig.~\ref{fig_3}). 

%\ali{[why you did not use spectrum fluctuation monitor in APs too in the single hop scenario, previous paragraph description?]}

Basically, cognitive radio devices will be deployed in order to achieve two goals. The first one is to relay the data of the heterogeneous networks and/or devices to the Internet and the second one is to provide connectivity to the disjoint networks i.e. non-CR devices.  In the latter case, when CRNs deployed with partially destroyed networks to provide connectivity, CR devices will first perform the discovery of the partially destroyed networks i.e. infrastructure discovery and then tune themselves to the appropriate operating frequency of the disjoint network. In this manner, CR device will help the disconnected network to restore their connectivity.

More precisely, when CR devices are deployed in an area, they first perform a neighborhood discovery. Thus, CR devices will discover still alive non-CR devices. In the meantime, non-CR devices will learn the presence of CR devices in the form of new neighbors. Note that this process can be completely transparent to non-CR devices and may not necessitate any reconfiguration.
One example is if a non-CR device after the disaster becomes disconnected from the whole network, with the arrival of one or more CR devices, the connectivity of that non-CR device can be re-established.
%Thus, every non-CR device shall choose its communication : with another non-CR device or to the closest CR device to reach to the sink or Internet.

\subsubsection{Cognitive Multi-Radio Mesh Routers (CMRs)}
\label{ap}

Cognitive multi-radio mesh routers will be deployed in fixed locations. The main responsibility of these cognitive multi-radio mesh routers is to perform inter-communication between cognitive radio devices and internet portal point to facilitate data transfer and connectivity to Global Internet.

These cognitive multi-radio mesh routers can operate in two scenarios: (1) single-hop, and (2) multi-hop. In the single-hop scenario, cognitive multi-radio mesh routers perform channel monitoring, keep track of the spectrum occupancy, generate spectrum opportunity map~\cite{arslan} and facilitate cognitive radio devices in their reliable channel selection decision. Therefore, in this context, cognitive multi-radio mesh routers are responsible for channel assignment of the CR devices too. In single-hop scenarios, it may happen that non-CR devices generate concurrent transmissions over the spectrum. Thus, cognitive multi-radio mesh routers have to monitor the spectrum, perform sensing to detect free channels, and implement a scheduling algorithm to regulate the transmissions from CR neighbors to cognitive multi-radio mesh routers in such channels, limiting thus the contention. A simple polling mechanism as the one proposed in the IEEE 802.11 standard can be used here to schedule
transmissions between CR's and cognitive multi-radio mesh routers.
Single-hop scenario can be further classified into standalone approach and coordinated approach, based on the way spectrum opportunities are generated and distributed, as explained hereafter.

In the multi-hop scenario, cognitive multi-radio mesh routers are responsible for data relaying, as well as they will also implement scheduling algorithm to order the transmissions among their CR neighbors, and to limit contention. %Cognitive radio devices are themselves responsible for their channel selection decision. % Cognitive multi-radio mesh routers will also implement scheduling algorithm to order the transmissions among their CR neighbors, and to limit contention.

%In both single-hop and multi-hop scenario, non-CR devices may be transmitting. Thus, access points have to monitor the spectrum and do scheduling to detected free channels. \\

%Moreover, in single-hop scenario, access points are responsible to perform channel monitoring and to help CR devices in selecting reliable channels to be used.

% \ali{[Mubashir, there is still problem in your classification... why AP are monitoring the spectrum in multi hop if they only relay messages? what about what we have discussed that in multi-hop or single-hop the APs have also to implement a scheduling algorithm to order the transmissions from CR neighbors to APs, and than limit contention? why in single hop AP have to monitor the spectrum? you should explain this. Is it because other PRs could be transmitting? so if the case, in multi-hop and single hop the APs have to monitor the spectrum and do scheduling at detected free channels. I can not see now how the APs in multi-hop will only relay messages...]}

%Mubashir, give reference of co-operative spectrum sensing with appropriate links and references....~\cite{arslan}
%occupancy of primary radio nodes~\cite{arslan}

%CR devices can communiate in ad-hoc manner to realy to APs or the gateways to the internet.

The monitoring of the spectrum and generation of spectrum opportunity map by the cognitive multi-radio mesh routers can be done in two fashions:

\begin{itemize}
\item Standalone Approach: In this approach, cognitive multi-radio mesh routers themselves monitor the spectrum fluctuations {\it without any feedback from or coordination with the CR devices}. Moreover, cognitive multi-radio mesh routers can coordinate with other cognitive multi-radio mesh routers to share spectrum monitoring related information. In this manner, cognitive multi-radio mesh routers maintain a database of the corresponding geographic area and the designated radio spectrum.

\item Coordinated Approach: In this approach, cognitive multi-radio mesh routers generate and maintains spectrum opportunity map {\it by getting the feedback from cognitive radio devices}. These cognitive radio devices that are disperse around the vicinity of cognitive multi-radio mesh routers detect radio spectrum activity and sends this information to the nearest cognitive multi-radio mesh routers. This information contains the channel id, channel utilization time, frequency of the channel, etc.

\end{itemize}

Irrespective of any approach adopted by the cognitive multi-radio mesh routers i.e. standalone or coordinated approach in single-hop context only, the collected information about spectrum occupancy will be first analyzed by {\it Spectrum Fluctuation Monitor}, which is then stored in Spectrum Opportunity Database. This database can then be used to make predictions about the spectrum utilization. Fig.~\ref{fig_3} shows the function of a Spectrum Fluctuation Monitor maintained over the cognitive multi-radio mesh routers. Then based upon this, an intelligent spectrum decision can be perceived by the cognitive multi-radio mesh routers and will be communicated to the cognitive radio devices to help them select reliable channels (cf. section~\ref{surf} for more details). Moreover, cognitive multi-radio mesh routers are also responsible to take care of the contention over the medium.

%\ali{[same comment here, this figure is very ugly compared to the new one. be more creative.]}

\subsubsection{Internet Portal Point}

Internet Portal Point are devices that serves as gateways to the Internet. These devices can be stationary or mobile; equipped with powerful communication medium, for e.g. satellite-link. They are responsible for sharing Internet bandwidth to, as well as gathering data from, cognitive multi-radio mesh routers and transfer it to the Internet. %\vspace{-0.4cm}

%Moreover, they are also responsible for sharing Internet bandwidth to cognitive multi-radio mesh routers. %In fact, by sharing the Internet bandwidth with cognitive multi-radio mesh routers, the coverage area of the partially destroyed network can be increased. \ali{[I still have problems with this sentence. what do you want to say here? the coverage of the destroyed network is extended to the whole internet? but this was done before the disaster too...]}\\

%In the proposed framework, Internet Portal Point are gateways to the internet. They gathers data from Access Points and transfer it to the internet. These devices can be stationary or mobile and equip with powerful communication medium, for instance satellite-link. Moreover, they are also responsible for sharing internet bandwidth to access points.  \\

\subsection{Working Principle}
Initially the network is deployed having a single internet portal point device. This device should be connected with the global Internet through the satellite link. In the vicinity of this internet portal point, fixed cognitive multi-radio mesh routers are deployed which are directly connected with the internet portal point. Internet portal point shares the Internet connection with these cognitive multi-radio mesh routers. In order to increase the coverage area and provide last-mile connectivity, more cognitive radio devices can be deployed in multi-hop fashion to reach to internet portal point via cognitive multi-radio mesh routers. Then, cognitive radio devices are deployed in such a manner that they co-ordinate with the partially destroyed network nodes and help them to restore their connectivity, provide Internet accessibility or relay their data to the global Internet.

% \ali{[AP in a multi hop?? this is strange... why not use CRs in a multi-hop, they probably are less expensive. Another issue, instead of APs why not use wireless mesh routers (WMRs)? than, there will be sense in talking about multihop WMR, they will create a static mesh network, and CRs will gravitate among them, also establishing or not multi hop paths.]}

In this manner, the cognitive radio devices first discover the partially destroyed existing infrastructure through spectrum sensing and resource discovery mechanisms, which we will discuss in section~\ref{deploy}. Once detected, they forward the data of the partially destroyed network to the nearest cognitive multi-radio mesh router which further relay it to the central internet portal point and finally this data reaches to the global Internet. %\ali{[attention verify if this was not already said]}

%and connect to Internet a higher number of

\begin{figure}[htbp]
    \begin{center}
    {
        \includegraphics[width=12cm, height=6.5cm]{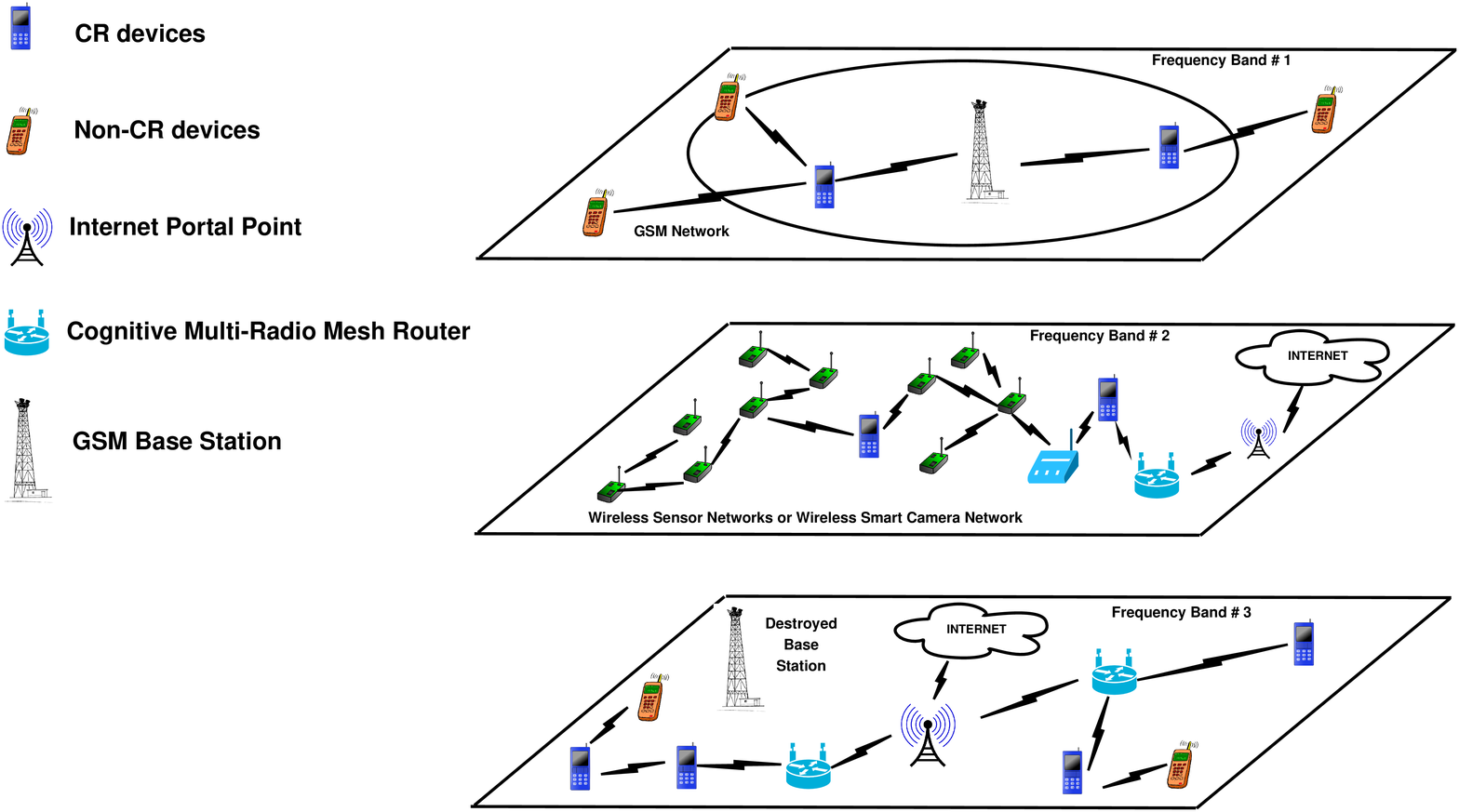} 
    }
\caption{Cognitive radio based internet access framework helps distinct network entities to restore their connectivity to the global Internet. }
\label{fig_13}
\end{center}
\end{figure}

\subsection{Application Scenarios and Advantages}
In a post disaster situation, one may have information from the partially destroyed existing infrastructures in order to have insights about the catastrophe area. For e.g., getting connected to a sink node of an smart camera network can play a vital role in rescuing lives in the area (cf. Fig.~\ref{fig_13}, the 2nd plane). 
%Refer this scenario to the Fig.~\ref{fig_13}, the 2nd plane.

%\ali{[this sentence is not understandable. rephrase it. what do you want to say here?]} 
%\ali{[again, not understandable? what do you want to say here?]}

%\ali{[Refer this scenario to the figure ~\ref{fig_13}, the 2nd plane. Put in the 2nd plane of the this figure, instead of Wireless Sensor Networks, put  Wireless Sensor Networks or Wireless Smart Camera networks.]}

%\ali{[ the goal of the figure ~\ref{fig_13} was to give a visual presentation of the applications you will describe here. Describe the DTN example, GSM also.]}

Another example in a post-disaster situation could be a static sink deployed in a wireless sensor network to collect the sensed information that may lose the connectivity to the Internet. And some how the collected data has to be delivered to a base station that will process it. In order to deliver this data to the base station, Internet connectivity is required. In addition, after the poster disaster situation, it might happen the reachability of humans to the region where the sink is located, might be hard to achieve or not be possible. Self-deploying mobile cognitive radio devices can facilitate this access by providing Internet connectivity to the sink.  %Fig.~\ref{fig_13}, the 2nd plane shows that a CR device is restoring the connectivity of static sink and upload the data to the Internet. 

%Fig.~\ref{fig_8} shows that a CR device is restoring the connectivity of static sink and upload the data to the Internet.

%\ali{[again why sink has a connectivity to the iNTernet... explain that some how the collected data has to be delivered to a base station that will process it, if this delivery is done through the internet, than you have this problem.]}

%\begin{figure}[htbp]
%    \begin{center}
%    {
%        \includegraphics[width=9cm, height=4cm]{./Figures/Sensor.eps} \vspace{-0.6cm}
%    }
%\caption{A CR device connecting the disjoint Wireless Sensor Network to the Global Internet.}
%\label{fig_8}
%\end{center}
%\end{figure}

An important advantage of the proposed architecture is that cognitive radio devices can be able to access real updates about weather forecast or seismic forecast of the post-disaster area from the natural disaster tracking system~\cite{noaa} and Geographical Information Systems (GIS)~\cite{esri}. Once they have this information, they can help rescue team members by guiding them towards safe passages and help them to reach to the nearest rescue base stations. Fig.~\ref{fig_13} shows different applications of the proposed framework, where it can restore the connectivity of distinct network entities and connect them to Global Internet.

%\ali{[the figure is great, but can still be improved. Which devices are the CRs? what is the bottom plane network (DTN?)? why don't you put the some cognitive mesh routers and the IPP? put a legend for the devices of your architecture.]}

%\ali{[there is a huge error in your figure .~\ref{fig_13} too. CRs can not be solated in the network, they have to be connected to another CR or to a mesh router... otherwise how they will communicate, they can not use non-CR nodes... so, your middle and bottom plane are wrong. in the bottom plane CRs have to pass through the non-CR to reach the AP?? correct this, add mesh routers and IPP]}

%\ali{[another suggestion, make the bottom plane in a way, it can be used as example for your simulation.]}

%\ali{[again, develop... once they have this information, how can they help to rescue?? doing what?]}

%Cognitive Multi-Radio Mesh Routers

%We now discuss some important issues and challenges regarding the connectivity and deployment of the proposed infrastructure. %\ali{[not true, you talk about applications.]}

\section{Deployment and Connectivity: Issues and Challenges}
\label{deploy}
We now discuss how the proposed framework address the issues and challenges concerning the deployment and connectivity of the CR devices to restore the connectivity of partially destroyed networks and to connect them to global Internet.

%\ali{[Mubashir, I have already commented this, stop saying "help a lot" this is an ugly way to say it, try to be more creative in your sentences "can add extreme value to the deployment...", or ``can be useful to the deployment'']}

\subsection{Network Deployment and Connectivity}
Depending on the application requirements, the deployment of cognitive radio devices may exhibit different network topologies: (1) single-hop (centralized), and (2) multi-hop (ad-hoc). Besides, traditional self-deployment techniques~\cite{eric} can add extreme value to the deployment of cognitive radio devices. 
When the cognitive radio devices are deployed in ad-hoc fashion and they create multi-hop network to reach the cognitive multi-radio mesh router, the main issue is regarding their connectivity.
In fact, cognitive radio devices must select reliable channels in order to ensure their connectivity. Otherwise, a CR device that intends to upload its data to the Internet may not be able to find relaying CR devices and a delay may occur which is undesirable in post disaster situations. Moreover, without any intelligent channel selection strategy, there will be contention and collisions, which results in packet losses. Thus, to deal with these issue, we propose to use our channel selection strategy SURF~\cite{tr} in conjunction with the proposed framework. SURF provides a good level of connectivity and is well suited for these scenarios (cf. Section~\ref{surf} for more details).

%In the deployed ad-hoc infrastructure, the main issue issue is regarding the connectivity of cognitive radio devices themselves. In fact, the cognitive radio devices should select reliable channels in order to ensure their connectivity~\cite{mubashir,tr}.

%\subsection{Network Deployment}
%Depending on the application requirements, the deployment of cognitive radio devices may exhibit different network topologies: (1) Ad-Hoc (2) Centralized, and (3) Hybrid. In ad-hoc environment, the cognitive radio devices build multi-hop network to reach to access point, while in centralized environment, the mobile cognitive radio devices are only allowed to communicate in a single-hop fashion with the access point. The hybrid is the mixture of two environments.

\subsection{Infrastructure Discovery}
Infrastructure discovery is another important aspect that need to be considered. Primarily because the deployment of the proposed framework depends on the knowledge of network that was operating previously. Insights such as whether the previously deployed network works with Base Stations (BS) or not, in ad-hoc mode or decentralized mode, with a sink, on which frequency, etc can aid the initial deployment and configuration of the proposed framework. Infrastructure discovery means the identification of the existing infrastructure, such as wireless sensor networks' nodes and sink, WiFi access points, GSM base stations, etc. This infrastructure discovery can be done by using advanced methods that relies on Software Defined Radio (SDR)~\cite{arslan} technology which scans the radio spectrum, search for beacons and radio signals, and identify the presence of any radio device.

In order to restore the connectivity of the partially deployed fixed telecommunication infrastructure like GSM base stations, prior knowledge about the deployed infrastructure is required. This will facilitate to analyze how much CR devices should be deployed and in which geographic region they should be deployed. The real challenge is in restoring the connectivity of networks where prior knowledge of the deployed infrastructure is not available. For instance, WLANs, that operates in the ISM band and can be heavily deployed in urban city regions and office buildings. %\ali{[not get your point here? there is a real challenge in restoring connectivity of networks that knowledge is not availabe and then you talk about WiFi?? what is the relationship with Wifi?]}
Thanks to the inherent capabilities of cognitive radio devices and methods such as~\cite{rob},~\cite{damljanovic}, the service discovery and identification of the available communication technology, such as Bluetooth, WiFi, is now possible. Moreover, these techniques can be also used to identify and distinguish different operating devices including Bluetooth piconets, individual Bluetooth devices, WiFi access points, sensor nodes, PDAs.  %Similarly, to identify the operating CR devices and to distinguish them with the partially destroyed network devices, bootstrapping protocol~\cite{jing} can also be applied. Thus, the aforementioned methods can be used to support the proposed framework.

%\ali{[I really recommend you to re-read this section many times in order to improve the writing. It is really necessary here. I tried, but not still good. Do it once you have had some rests. otherwise you will not be able to do it well]}

\subsection{Inter-network Coordination}
Inter-network coordination means how the cognitive radio devices communicate with distinct network entities. This challenge is addressed in the proposed framework by exploiting the inherent capabilities of CR devices. For instance, during the infrastructure discovery phase, CR devices already aware of distinct network entities and their operating frequency. Thus, to communicate with a distinct network entity, for example, a sensor node, a cognitive radio device will tune itself to the sensor nodes' operating frequency. In this manner, cognitive radio device have to select one channel to communicate with sensors and other channel to communicate with cognitive radio devices or cognitive multi-radio mesh routers.

\section{Channel Selection Strategy SURF for CR Devices and CMRs} % and Cognitive Multi-radio Mesh Routers}
\label{surf}

%Talk about simulations of your framework with SURF. You deployed CRs and disseminate massages blabla... Think how to include
%AP in the simulations. I advise you to put some APs in the simulations and see How many are reached, change the number of APs and redo
%simulations. You have to be more creative, these results are important and should prove the feasibility and usability of the framework.

When CR devices wants to upload their data to the Internet, they are required to communicate with the cognitive multi-radio mesh router over a particular channel. Without any intelligent channel selection strategy, concentration of all the cognitive radio devices over a particular channel could lead to contention and collision problems, which further reduces the connectivity to the global Internet. Thus, channel selection plays a vital role in efficient and reliable data relaying to the cognitive multi-radio mesh router and/or between to cognitive radio devices which operates in multi-hop fashion to reach the cognitive multi-radio mesh router.

In scenarios where cognitive multi-radio mesh routers are responsible for channel assignment to CR devices, a channel selection strategy proposed by us~\cite{tr,mubashir_wowmom} could be executed by the cognitive multi-radio mesh routers. Here, the mode of communication between the cognitive radio devices and cognitive multi-radio mesh routers is single-hop. But, when cognitive radio devices communicate with other cognitive radio devices and create a multi-hop network to reach to cognitive multi-radio mesh routers, the task of relaying data to the cognitive multi-radio mesh routers will be much more challenging. This is due to the diversity in the number of available channels and lack of any centralized authority. Therefore, keeping these challenges in mind, more recently we proposed a channel assortment strategy SURF~\cite{tr,mubashir_wowmom}. SURF channel assortment is performed in a distributed way and is based only on information locally inferred by CR devices. %\ali{[Why don't you say they are your work??it seems you are copying and it is not the case. ]} %In fact, by implementing the same strategy at the sender and the receiver, SURF helps both of them tune to the appropriate channel for
%undergoing transmissions or reception without the need of any prior information exchange or synchronization.

\subsection{Performance Evaluation}
To evaluate the feasibility and usability of the framework, we use SURF~\cite{tr},~\cite{mubashir_wowmom} to be implemented by CR devices to relay data towards cognitive multi-radio mesh router. %We now describe the simulation environment. 

%The same channel selection strategy SURF is also implemented by access points for overhearing purpose.

\begin{figure}[htbp]
    \begin{center}
    {
        \includegraphics[width=9cm, height=5.5cm]{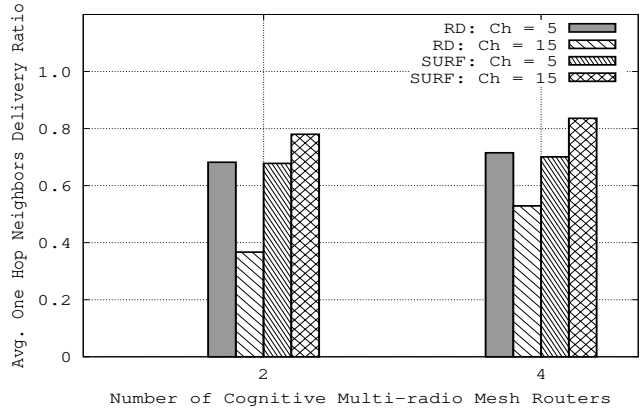} 
    }
\caption{Comparison of average one-hop neighbors delivery ratio for random (RD) and our strategy (SURF), when {\it channels=5} and {\it channels=15}, for varying number of cognitive multi-radio mesh routers.} 
\label{fig_12}
\end{center}
\end{figure}

\subsubsection{Simulation Environment}
%\ali{[what is missing here is a small disaster scenario, where you describe the CR deployment and the communication of a non-CR message to CRs...]}

We consider a post-disaster scenario in which the existing infrastructure is partially destroyed but still there are some alive non-CR devices that need Internet connectivity to deliver emergency messages, alerts or wants to deliver their data to the Internet (cf. bottom plane of Fig.\ref{fig_13}). In order to restore the connectivity and to help the rescue team members, we deploy an ad-hoc cognitive radio network in which CR devices communicate in multi-hop fashion in order to reach the nearest cognitive multi-radio mesh router. These cognitive mesh routers are deployed randomly and upload this data to the Internet via internet portal point. % Cognitive multi-radio mesh routers are randomly deployed in the network and the transmission range is set to 250m. %\ali{[here you could refer to the bottom plane in figure 5, but for this you have to accordingly modify the figure 5.]}

First, the channel selection phase starts in which CR devices execute SURF~\cite{tr,mubashir_wowmom} and select the best channel for transmission and/or overhearing. The message dissemination phase then starts, in which a randomly selected CR device, who received the data from the non-CR device, disseminates the message on the selected channel by setting the TTL. CR neighbor devices that are on the same selected channel will overhear the message, decrease TTL, redo the spectrum sensing, select the best available channel, and disseminate the message to the next-hop neighbors until TTL=0. In this manner, the message is traversed throughout the network and if at any stage, CR device and cognitive multi-radio mesh router are on the same channel, the message will be delivered to the cognitive multi-radio mesh router. 

%We then count that how many access points are reached. \\

%The graph will be look like this: x-axis (Total number of APs deployed) y-axis (No. of APs reached)

\subsubsection{Reachability of Cognitive Multi-radio Mesh Routers}
%In order to evaluate the performance of SURF, we compare it with an intuitive random
%strategy (RD) and the selective broadcasting strategy (SB)~\cite{agrawal}.

The data will be delivered to the Internet {\it iff} it is received by the one-hop neighbors of the cognitive multi-radio mesh router. The hop-by-hop delivery of this data is more difficult in the presence of PR nodes and multiple channels and it is difficult to guarantee that CR devices will have receivers when they transmit. %Primarily because under these constraints, it is not sure that CR devices will have receivers when they transmit. 
Thus, in order to check the data is successfully delivered to one-hop neighbors of cognitive multi-radio mesh router, we calculate the average delivery ratio of message reached to one-hop neighbors of CMRs. 

%\ali{[explain why this task is not trivial, having PRs and multi-channels is not sure CRs will have receivers when they transmit, bla bla bla...]} 

Fig.~\ref{fig_12} compares the average one-hop neighbors delivery ratio for random (RD) and our strategy (SURF), when {\it channels=5} and {\it channels=15}, for varying number of cognitive multi-radio mesh routers. Note that in RD strategy, channels are randomly selected to be used by CR devices for transmission
and/or overhearing, i.e. without any consideration to the ongoing PR and CR
activity over these channels. SURF allows the average delivery ratio of 65\% and 80\% messages to the one-hop neighbors of cognitive multi-radio mesh routers in the network, compared to 65\% and 40\% in the RD case, for {\it Ch=5} and {\it Ch=15} respectively. When there is a increase in the number of channels i.e. {\it Ch=15}, SURF allows the average delivery ratio of 80\% messages to the one-hop neighbors of cognitive multi-radio mesh routers in the network, compared to 40\% in the RD case. In fact, this happens since during
channel selection SURF considers both the PR nodes and number of CR neighbor
receivers. These results show the good level of network connectivity provided by SURF, suitable for reliable data relaying.

\section{Conclusion}
\label{sec:conclusion}
In this paper, we have discussed that cognitive radio network can be served as a core technology which enables partially destroyed networks to restore their connectivity. In this regard, we presented a cognitive radio based Internet access framework for disaster response network deployment in challenged environments. This architecture is specially designed to cater the needs of challenged environments. We then highlighted issues and challenges in the deployment of this architecture. We believe that this work can be served as a basis to build new algorithms and protocols that lead us to federate various communication paradigms. 

\bibliographystyle{plain}
\bibliography{IAFCRN}

\tableofcontents
\end{document}